\documentclass[prl,twocolumn,showpacs]{revtex4-1}

\usepackage{graphicx}
\usepackage{ulem}
\usepackage{dcolumn}
\usepackage{bm}
\usepackage{natbib}

\begin{document}

\title{A New Cosmological Model for the Visible Universe and its Implications}

\author{Branislav Vlahovic}
\email{vlahovic@nccu.edu}
\affiliation{Department  of Physics, North Carolina Central University, 1801 Fayetteville 	Street, Durham, NC 27707 USA.}


\begin{abstract}
\noindent Assuming that the universe is homogenous and isotropic and applying Birkhoff theorem or Gauss's flux theorem for gravity, it follows that the gravitational field of the visible universe on the imaginary Gauss's surface can be calculated as if the entire mass of the visible universe is located at one point. Taking into account that the mass of the visible universe is about $M$ = 2x$10^{53}$ kg, it appears that the entire visible universe is inside a {\it photon sphere} of radius $R_{ps} = 14.3$ Gpc. The current model for the visible universe must be corrected to account for the fact that the measured horizon distance of 14.0 $\pm$ 0.2 Gpc is not a straight line. Rather it is an arc of a circle with that length, because all photons are forced to follow circular orbits, since they are inside the photon sphere. Our model interprets the visible universe as the surface of a sphere (or the inside of a spherical shell) with radius 4.46 $\pm$ 0.06 Gpc and an event horizon, located on that sphere (shell), with size of 14.0 $\pm$ 0.2 Gpc. The model predicts the redshift of the cosmic microwave background (CMB) and time dilatation of Type Ia supernovae by gravitation. It explains,
without inflation theory, the isotropy and uniformity of the CMB.   It predicts the correct value for the Hubble constant $H_0$ = 67.26 $\pm$ 0.90 km/s/Mpc, the cosmic expansion rate $H(z)$ in agreement with observations, and the speed of the event horizon.  Through relativistic energy correction,  model also provides an explanation for the critical density without recourse to dark matter. It explains that type Ia supernovae redshifts are not related to the accelerated expansion of the universe and dark energy. It explains the reason for the established discrepancy between the non-covariant version of the holographic principle and the calculated dimensionless entropy $(S/k)$ for the visible universe, which exceeds the entropy of a black hole.   The model is in agreement with the distribution of radio sources in space, type Ia data, and data from the Hubble Ultra Deep Field optical and near-infrared survey.
\end{abstract}

\maketitle

\section{Justification for a New Model}

 The observable universe is defined as a sphere, centered on the observer and from our perspective it appears that the radius is $R_0$ = 14.0 $\pm$ 0.2 Gpc (about 45.7 Gly).  The value $R_0$ is the particle horizon and the quoted result corresponds to the direct WMAP7 measurements and the recombination redshift $z$ = 1090 $\pm$ 1 \cite{2}.

A problem with this model is that it does not take into account bending of light by mass. It assumes that light is expanding straight, radially, in all directions for 14 Gpc. As it is well known if a photon passes a massive object at an impact parameter $b$, the local curvature of space-time will cause the photon to be deflected by an angle
 \begin{equation} \label{mass__1}
 \alpha = \frac{4GM}{c^2b}.
\end{equation}
If photons are in a region of space where gravity is sufficiently strong, a {\it photon sphere} of radius
\begin{equation} \label{photon__1}
 R_{ps} = \frac{3GM}{c^2},
\end{equation}
then the photons will be forced to travel in orbits. It is usually stated that the photon spheres can only exist in the space surrounding an extremely compact object, such as a black hole or a neutron star. However, as it will be shown the concept is also applicable to the visible universe.

Assuming the basic principles of cosmology, that the universe is homogenous and isotropic (uniformity of mass density $\rho$) and applying Gauss's flux theorem for gravity, it follows that
 \begin{equation} \label{Gaus__1}
 \oint {\bf\Phi} d{\bf a} = -4\pi GM.
\end{equation}
The gravitational field {$\bf\Phi$} on the boundary of the imaginary sphere that surrounds mass $M$ is exactly the same as it would have been if all the mass had been concentrated at the center of the sphere.

 This means (or by applying Birkhoff's theorem) that we can assume that the entire mass of the visible universe (considering the visible unverse as a sphere) is located in the center of that sphere. Using for $M$ = $10^{23}M_{\odot}$ = 2x$10^{53}$ kg, it gives $R_{ps}$ = 14.3 Gpc. All photons that are inside sphere of 14.3 Gpc will be forced to follow circular orbits. Since the visible universe is smaller or about the calculated $R_{ps}$ value all photons will be affected. Therefore, when we speak about the size and the radius of the visible universes we must take into account the bending of light. We cannot say that the visible universe is a sphere with a radius of 14 Gpc since photons cannot travel straight. The measured horizon distance of 14 Gpc is not the length of a straight line, because of the bending of light it is an arc of a circle with a length of 14 Gpc. The radius of that circle is $14.0/\pi=4.46$ Gpc. This is an important result that will have as we will see significant implications.

\begin{figure}[h]
		\includegraphics[width=7.5cm]{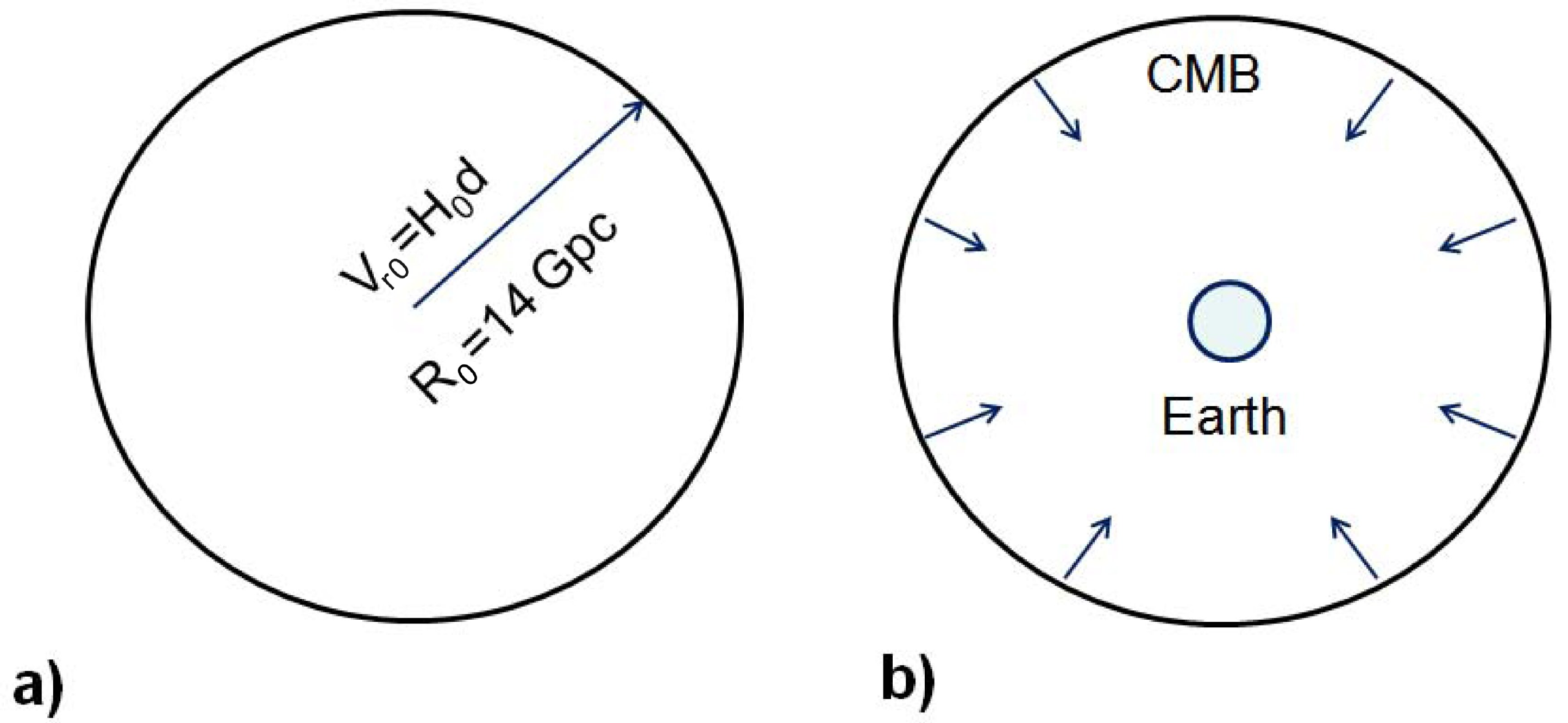}
		\caption{\label{fig1}
a) The visible universe and expansion of space and b) the current interpretation of the CMB as visible from the Earth.
}
\end{figure}

The inconsistency between the current and our proposed model is that according to the current model the visible universe is a sphere. While our model requires that our galaxy and other galaxies are inside a thin three-dimensional shell, not inside the sphere.  In addition, if the boundary of the visible universe approximately corresponds to the physical boundary of the universe at the present time (if such a boundary exists) the current model implies that Earth is exactly at the center of the visible universe. This is a non-Copernican model, as shown in Fig. \ref{fig1}, that is difficult to accept, 
since this model places us at the center of the universe, whereas according to our model we and our galaxy should be presented as a point or a localized area on the surface of the sphere (shell) that represents the universe.

It is important to note that proposed thin shell model is not in a conflict with observations, that we do not see the edge of the universe. We cannot  observe it because of the bending of light.

To be consistent with general relativity we will consider a model of the universe that is based on the assumption that the presence of matter or energy causes warping, or curvature of spacetime. 



 We will consider a three dimensional model in which the universe is an expanding thin shell with thickness much smaller than its radius. Such model can be justified by the short time interval of the Big Bang universe creation and its further expansion. We may assume, as it has been always emphasized, that galaxies do not move through space and that the universe is not expanding into empty space around it, for space does not exist apart from the universe. There is no other space than that associated with the shell.
 The motions of all galaxies and propagation of the light are confined to the volume of the shell, which expands with a radial velocity.

  The dynamics of thin shell models has been investigated earlier. It was first introduced by Israel \cite{Israel}, in the framework of the special-relativity by \cite{Czachor}, and a systematic study in the framework of general relativity is done for instance in  \cite{Berezin} and \cite{Krisch}.  However, our focus will be very different. We will present significant implications of the shell model when combined with a new interpretation of experimental data. The new combined model that will be considered satisfies the basic assumptions that the universe is isotropic and uniform. The isotropy of the model will be explicitly demonstrated when the uniformity of the CMB will be considered.

\section{Tests and Implications of the New Model }

We will now apply our model to further elaborate it and to test its two main outcomes, for a factor of $\pi$ smaller predicted size of the visible universe, and requirement to take into account the total mass inside the Gauss's imaginary sphere when a gravity field is estimated.

Let us first use the model to explain CMB redshift as the gravitational redshift of light. It is one of the central predictions of metric theories of gravity, such as general relativity, that photons will lose energy leaving a massive object and gain it when moving toward a gravitational source. This is experimentally verified first in \cite{2a} and more recently in \cite{2b}.

The redshift between two identical frequency standards placed at rest at different strengths in a static gravitational field is:

 \begin{equation} \label{redshift__1}
  \Delta\nu/\nu = -\Delta\lambda/\lambda = \frac {\Delta U} {c^2} = \frac {GM} {Rc^2} - \frac {GM} {R_0c^2},
  \end{equation}

where $R$ is the size of the universe at time of decoupling obtained by scaling $R_0$ = 4.46 Gpc by the redshift $z$=1090. Using the value for mass of visible universe $M$ as above, we obtain for the gravitational redshift  1166, which is in agreement with observed value, taking into account uncertainty in the $M$.

\begin{figure}[h]
		\includegraphics[width=7.5cm]{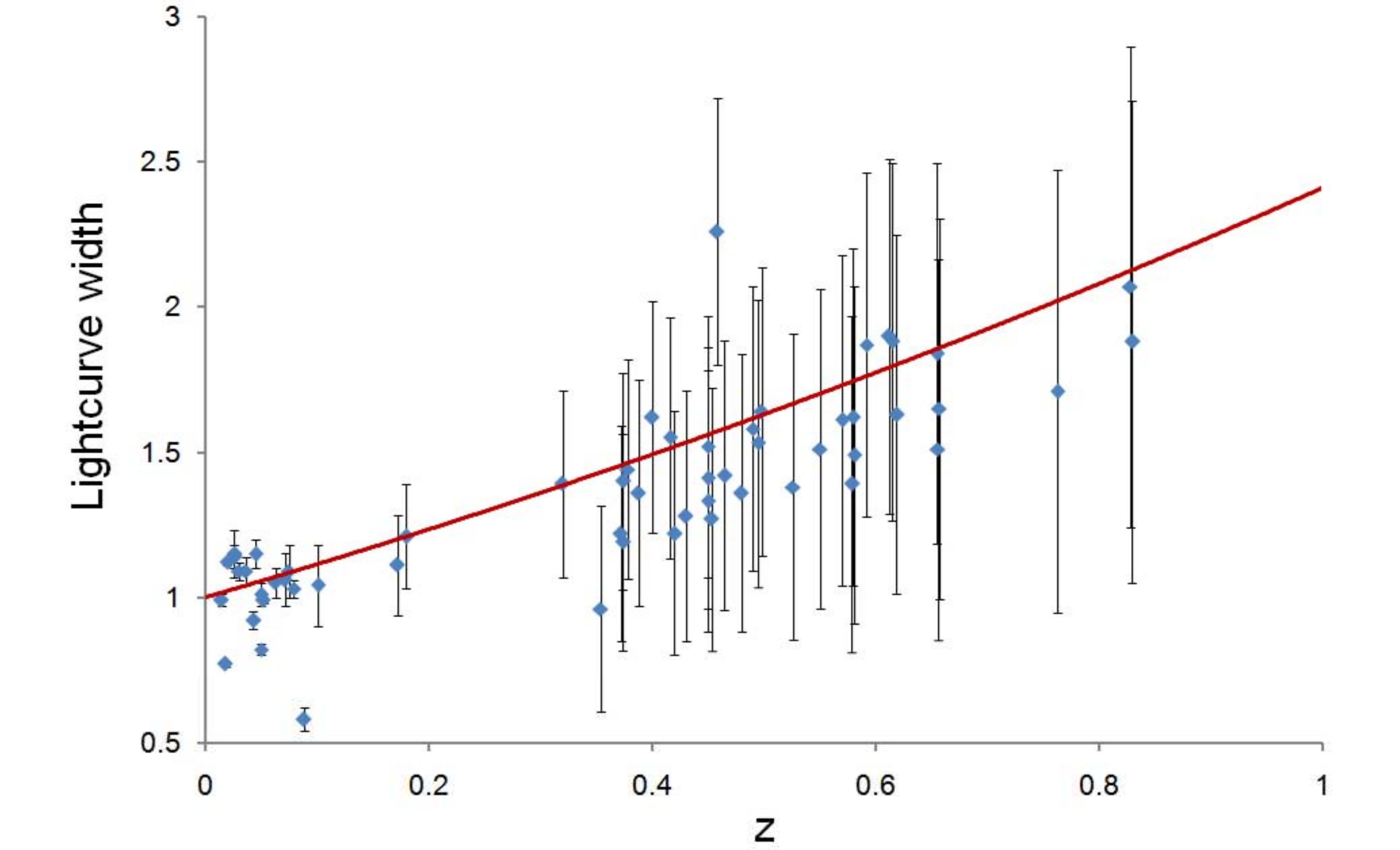}
		\caption{\label{fig-time}
The time dilatation of Ia supernovae. Solid line are calculations for different z using equation (\ref{time}), with distances corrected for factor $\pi$ in accordance with our model. Data points are from \cite{2c}.
}
\end{figure}

The approach could be further tested by calculating time dilatation of Type Ia supernovae (SNe Ia), because a clock in a gravitational potential U will run more slowly by a factor $\Delta t$

\begin{equation} \label{time}
  \Delta t  = 1+U/c^2,
  \end{equation}

 as compared to a similar clock outside the potential. Calculating for $z$=0.5 gives factor of 1.63 and for $z$=1 factor of 2.41, which is inside of one standard deviation with results \cite{2c}, see Fig. \ref{fig-time}.

Slightly higher results for the CMB redshift and for time dilatation than observed are obtained because in our calculations we used in (\ref{redshift__1}) and (\ref{time}) for $M$ the total mass of the universe, as we are on the top of the shell, while we are somewhere inside the shell.

\begin{figure}
		\includegraphics[width=8cm]{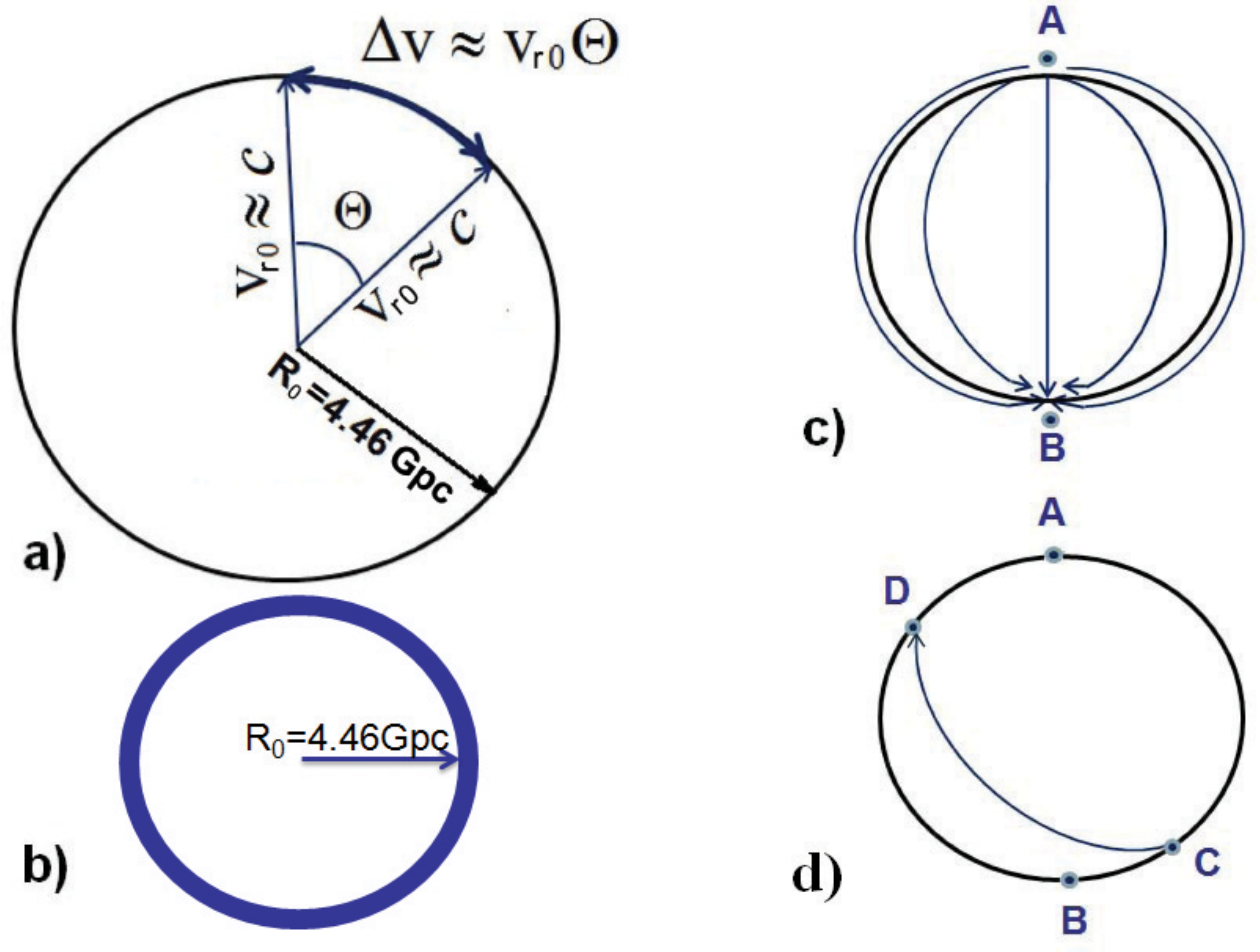}
		\caption{\label{fig2}
a) The visible universe as a surface of the sphere with radius $R_0$ = 4.46 Gpc that expands with speed close to the speed of light, b) The visible universe as an expanding shell with thickness much smaller than radius, c) Observable universe, as seen by an observer from the point A, is a surface of a sphere, with event horizon located in the point B, d) CMB visible from Earth (by observer in point A) is originated in point B and CMB visible from another place in the universe (point C) is emitted in the point D.
}
\end{figure}

Let us now introduce our model for the visible universe that is in accordance with the general relativity, bending of space and light. In our model the observable universe is the largest visible area (from the point of the observer) on the surface, Fig. \ref{fig2}a (inside of the shell, Fig. \ref{fig2}b), of the sphere that represents the universe. So, by this definition the particle horizon distance will be the largest possible distance on that surface (shell). If for instance the universe is the same size as the observable universe and an observer is located on the north pole of that sphere at point A, the particle horizon for that observer will be the point B on the south pole of the sphere and the observable universe will be the entire surface (shell) of the sphere. This is shown in Fig. \ref{fig2}c.

So can we, in addition to the statement that the current model is not taking into account the general relativity bending of light, prove which model of the observable universe is correct?

 First let us observe that the current and our model predicts significantly different sizes and different volumes of the visible universe. The current model claims that the visible universe is a sphere with radius $R_0$ = 14.0 $\pm$ 0.2 Gpc, and our model predicts that it is a sphere with a circumference of 28.0 $\pm$ 0.4 Gpc, or with a radius $R_0$ = 4.46 $\pm$ 0.06 Gpc, Fig. \ref{fig2}a.

We will now prove that the model that predicts the larger size for the visible universe is not in agreement with observations. As it is shown above the larger size gives wrong predictions for the CMB redshift and time dilatations at a particular $z$, for factor of $\pi$. In addition, we will now prove that the current model is also not in accordance with the holographic principle, which should be fulfilled for any successful model. The entropy of the visible universe is calculated in \cite{3} and it is shown that the dimensionless entropy $S/k$ is 8.85 $\pm$ 0.37 times larger than allowed by a simplified and non-covariant version of the holographic principle, which requires that the entropy cannot exceed that of a black hole.

It was argued in \cite{3} that by the holographic principle the entropy $S/k$ has an upper limit equal to that of a black hole:
\begin{equation} \label{GrindEQ__1}
\left(\frac{S}{k} \right)_{Uni} \le \left(\frac{S}{k} \right)_{BH} =\frac{4\pi R_{S}^{2} }{l_{P}^{2} }
\end{equation}
where $\left(\frac{S}{k} \right)_{Uni} $ is the entropy of the visible universe, $\left(\frac{S}{k} \right)_{BH} $ is the entropy of a black hole, $l_{P}^{} $ is the Planck length, and $R_{S}^{} $ is the Schwarzschild radius $R_{S}^{}$ = 2 $GM$.

Equation (\ref{GrindEQ__1}) requires
\begin{equation} \label{GrindEQ__2}
\frac{\left(\frac{S}{k} \right)_{Uni} }{\left(\frac{S}{k} \right)_{BH} } =R_{BET}^{4} \le 1
\end{equation}
where $R_{BET}$ is the Bond, Efstathiou, and Tegmark dimensionless shift parameter \cite{3a} defined as
\begin{equation} \label{GrindEQ__3}
                R_{BET} =\frac{\sqrt{\Omega _{m} H_{0}^{2} } }{c} R_0
\end{equation}
Taking from \cite{2} the size for the radius of the visible universe as $R_0$ = 14.0 $\pm$ 0.2 Gpc gives the value $R_{BET}$= 1.725 $\pm$ 0.018, and hence
\begin{equation} \label{GrindEQ__4}
                             R_{BET}^{4} = 8.85 \pm 0.1,
\end{equation}
which, as pointed out in \cite{3}, is in contradiction with equation (\ref{GrindEQ__2}) for 21$\sigma$. Therefore the current (larger) model for the visible universe presented by figure \ref{fig1} violates the holographic principle. However, the author of \cite{3} at this point speculates that equation (\ref{GrindEQ__2}) was fulfilled in the past when the radius of the universe was
\begin{equation} \label{GrindEQ__5}
                                  R \le 8.4 \pm  0.1~Gpc
\end{equation}
and further speculates that is when the cosmic deceleration ended and  acceleration began.

Both of these assumptions are based on the evaluation of the relation (\ref{GrindEQ__3}) by using the present time Hubble constant $H_0$ instead of the cosmic expansion rate $H(z)$ that corresponds to the size of the universe at that time. It is well known that the expansion rate $H$ is function of redshift $z$, and that it increases significantly with $z$. For instance, using combined Sn Ia, WMAP3, and SDSS data, it was concluded in \cite{4} that $H(0.5)/H_0 = 1.3 \pm  0.1$, $H(1.0)/H_0 = 1.8 \pm  0.2$, and $H(1.4)/H_0 = 2.4 \pm  0.4$. Therefore the radius of the visible universe at past times cannot be obtained from relation (\ref{GrindEQ__3}) by using $H_0$.
As we will show later, our model predicts that the product of the cosmic expansion rate $H(z)$ and the radius of the universe $R$ is a constant with a value equal to the expansion speed of the universe, see equation (\ref{GrindEQ__10}).

According to our model of the observable universe
\begin{equation} \label{GrindEQ__6}
                                R_0 = 4.46 \pm 0.06~Gpc
\end{equation}
\noindent the ratio in equation (\ref{GrindEQ__2}) is satisfied and it has been always $\le 1$ as it is required by the holographic principle.  Equation (\ref{GrindEQ__3}) is actually another expression for the equation (\ref{GrindEQ__10}), which can be seen by putting in equation (\ref{GrindEQ__3}), $\Omega_m$=1. Therefore, because equation (\ref{GrindEQ__3}) and (\ref{GrindEQ__10}) are the same equations, and because in equation (\ref{GrindEQ__10}) speed $v_{r} \le c$, the inequality (\ref{GrindEQ__2}) must be always $\le 1$ as it is required by the holographic principle.

In addition, let us mention here that the smaller value for the radius of the universe allows for an explanation of the universe without inflation theory and superluminal speeds. Assuming a value for the present age of the universe $t_0$ = 13.75$ \pm$ 0.17 Gy \cite{2} and that the universe expanded with the constant speed close to $c$, gives
\begin{equation} \label{GrindEQ__7}
R_0 = ct_0 = 4.22 \pm 0.06~Gpc  ,
\end{equation}
which is in agreement with our model.

In our model, space is expanding from the moment of creation (Big Bang) with speed close to the speed of light. As mentioned earlier, light is confined to the surface (shell) and can only travel on the surface (shell) of the sphere; for that reason we cannot point to the center of the universe. As seen from our galaxy, all other galaxies are moving away from us (and from each other) with the speed $v = v_{r0}\Theta$ (where $v_{r0}$ is radial speed of expansion and $\Theta$ is azimuthal angle, Fig. \ref{fig2}a), which is actually the Hubble law $v = H_0 \times distance$.  Taking into account the age of the universe $t_0$ and that $v_{r0}$ is close to $c$, it gives the radius of the sphere $R_0$ = 13.75 $\pm$ 0.17 Gly and a distance from A to B (Fig. \ref{fig2}c), which represents in comoving distances the size of the observable universe, that is equal to about 43 Gly. Using
\begin{equation} \label{GrindEQ__8}
v = v_{r0}\Theta = H_0R_0\Theta
\end{equation}
and expressing $R_0$ in Mpc, it is easy to calculate a value for the Hubble's constant $H_0$ = 71.17 $\pm$ 0.86 km/s/Mpc,
which is in agreement with the experimental data.

Setting into (\ref{GrindEQ__8}) the radius $R_0$ = 4.46 $\pm$ 0.06 Gpc, which is the value obtained by our model using for the particle horizon 14.0 $\pm$ 0.2 Gpc, gives
 \begin{equation} \label{GrindEQ__9}
H_0 = 67.26 \pm 0.90~km/s/Mpc,
\end{equation}
which is also in agreement with the experimental data. We will further use this value, since the age of the universe is model dependent, while the value for the particle horizon is obtained by direct WMAP7 measurement, without needing the details of the expansion history \cite{2}. Let us here note that the current model with $R_0$ = 14 Gpc will give wrong estimate for $H_0$ by a factor $\pi$.

\begin{figure}[h]
		\includegraphics[width=7.5cm]{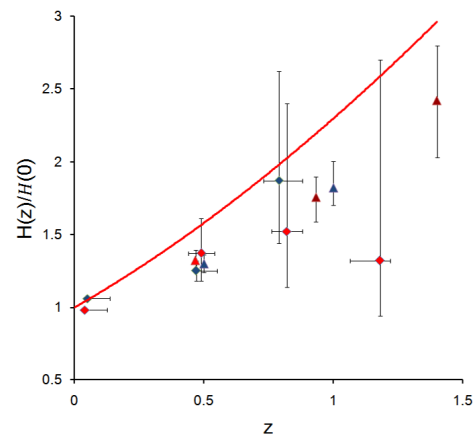}
		\caption{\label{Fig-H_z}
The solid line are calculations for $H(z)/H_0$ at different z using equation (\ref{GrindEQ__10}), with distances corrected for a factor $\pi$, in accordance with our model. Data points are from \cite{4}.
}
\end{figure}

However, let us here rewrite equation (\ref{GrindEQ__8}) in the form
\begin{equation} \label{GrindEQ__10}
                      v_r= H(z)R(z)
\end{equation}
to emphasize that the product $HR$ is a constant that has a value equal to the speed of radial expansion. From this equation one can also see that the cosmic expansion rate $H(z)$ changed with time and that it must have been larger for an earlier universe, if the universe expanded at an approximately constant speed. Equation (\ref{GrindEQ__10}) gives for $H(0.5)/H_0 = 1.6$, $H(1.0)/H_0 = 2.3$, and $H(1.4)/H_0 = 3.0$ which agrees with \cite{4} within three standard deviations, Fig. \ref{Fig-H_z}.

Equation (\ref{GrindEQ__10}) is basically the same as the first Friedmann equation
\begin{equation} \label{GrindEQ__11}
                      H^2= \left(\frac{\dot{a}}{a} \right)^2
\end{equation}
where for a closed 3-sphere universe the scale factor $a$ corresponds to the radius of curvature of the universe.  As shown above, our model predicts proper value for $H(z)$, while the current model is off by a factor of $\pi$. It is important to note that both the Hubble law and the first Friedmann equation follow from our model.

Let us here mention that uniform expansion of space without inflation theory and with subluminal speeds, which are assumptions of our model, is not in contradiction with uniformity of the CMB. Rather, as we will show below, the uniformity of the CMB is actually a natural consequence of our model.

The uniformity of the CMB and the general sameness of the observable universe at very large scales are explained by the inflation model, which assumes that a tiny region much smaller than the nucleus of an atom expanded to become the entire part of the observable universe. By the inflation model, when we are observing CMB on the two opposite sides of the universe, we are looking at parts of the universe that were identical when inflation began.  For that reason the CMB emitted by these two parts must be the same. However, let us here note that inflation ended when the universe was only $10^{-32}$ seconds old. From that time to the era of decoupling, the universe went through a period of the most significant transformations in which the density changed from $10^{38}$~kg/m$^3$ to $10^{-17}$~kg/m$^3$ and temperature changed from $10^{29}$~K to about 3000~K. The high degree of isotropy observed in the microwave background indicates that any density variations from one region of space to another at the time of decoupling must have been small, at most a few parts in $10^5$ and that the temperature variations between any two places are smaller than 30-40 millionths of a Kelvin from place to place in the sky. Keeping in mind that during this period of transformation some regions of the universe were separated by as much as 36 Mly and were not able to communicate for a period of 380,000 years, it is at least surprising that all of them would end up with almost exactly the same density and temperature. It is also important to note that this almost perfect uniformity in CMB cannot explain the formation of larger structures such as galaxies and clusters without introducing cold dark mater. In addition, Gurzadyan and Penrose recently have identified, within the CMB data, families of concentric circles over which the temperature variance is anomalously low \cite{GP}. The existence of these circles would not be easily explained within standard inflationary cosmology. We will show that there is another possible solution to explain uniformity of CMB, without inflation theory, dark matter, and dark energy if our model is applied.

In our model, looking from the position of our galaxy (marked by A), the place of decoupling (the surface of last scattering) is on the opposite side of the sphere (marked by B), Fig. \ref{fig2}c. Regardless of the direction we chose to measure CMB (for instance from point A looking in any arbitrary chosen direction), we will always measure CMB at the point which is on the opposite side of the sphere (in this example point B). The reason is that the length of the arc on the sphere represents distance, which also represents past time. The point that is at the largest distance away from point A is the point B, which represents the surface of last scatter, since we cannot see beyond that distance.  It is important to notice that measuring the same CMB by looking in the opposite directions of the universe does not represent or reflect the uniformity of the universe at the time of decoupling, because we always measure CMB originated from the same point regardless of the direction of observation.  For that reason we always must obtain the same result. If from point A we observe to the right, left, backward or forward we will always measure CMB originated from the point B. Small variations for the CMB are possible and they are observed, but they are the result of the interaction between matter and light during its travel. For instance, depending on the direction we choose to measure CMB, light will travel from point B to A through different galaxies and will interact with different amounts of matter, which may result in the small observed variations of CMB. The observed fluctuations in the CMB are therefore created as the photons pass through nearby large scale structures, a phenomenon known as the integrated Sachs-Wolfe effect. The correlation between the fluctuations in the CMB and the matter distribution is well established \cite{4a}\cite{4b}\cite{4c}\cite{4d}.

To establish a connection between the uniformity of the earlier universe at the time of decoupling and the CMB we will need to make a completely different kind of measurements of the CMB. We can see the CMB in any direction we can look in the sky. However, we must keep in mind that the CMB emitted by the matter that would ultimately form for instance the Milky Way is long gone. It left our part of the universe at the speed of light billions of years ago and now forms the CMB for observers in remote parts of the universe, actually exactly for an observer at the point B. For instance, if we perform measurement of the CMB at the point C, we will measure the CMB emitted by matter at the point D, Fig. \ref{fig2}d.  To measure uniformity of the universe at the time of decoupling we will need to measure the CMB in at least two different points on the sphere. If, for instance, the measurements from points A and C give the same result, then and only then may we speak about the uniformity of the universe at the time of decoupling. However, such measurements are not possible at the present time.

Let us here note that our model also may predict the families of concentric circles with anomalously low temperature variance, found in \cite{GP}. Since observed CMB is emitted at one point, but it is arriving to us from different directions, the interference may in principle create the observed patterns.

The presented model has significant consequences for current cosmological theories. It explains uniformity in the CMB without inflation theory. The model also removes any superluminal speed, since all galaxies in the model are by definition moving with speeds less than or equal to the speed of light. However, the size of the observable universe or the radius of the particle horizon is $\pi ct_0$.
 This is between the values $3cH_0^{-1}$ = $3ct_0$, which corresponds to the particle horizon in Einstein-de Sitter universe ($\Omega_m$ = 1 and $\Omega_\Lambda $ = 0) and $3.4ct_0$, which corresponds to the currently favored cosmological model $\Omega_m$ = 0.3 and $\Omega_\Lambda $ = 0.7  \cite{5}. The velocity of the particle horizon of this model is $2c$ which is the same as in the Einstein-de Sitter model.

It is important to note that up to this point our model assumed that space is expanding. However, because it describes the universe and all observable without inflation and superluminal speed, it allows us to consider the possibility that the expansion of the space is result of radial motion of the galaxies with speed close to the speed of light.

So, let us consider here a model with galaxies expanding from the center of the universe (place of Big Bang) with a radial speed (as it will be shown later) close to the speed of light. All galaxies are on the surface of the sphere (inside the shell). The motion of galaxies with the radial speed close to the speed of light, incorporated in this model, provides a valid reason for applying special relativity. For instance, this allows to introduce an interesting new idea for the interpretation of the missing mass, dark matter, and dark energy. The current assumption is that the universe contains 4\% of matter, 23\% of dark matter, and 73\% of dark energy. However, if galaxies are moving with the speeds close to $c$, we should take into account the increase of the energy due to this relativistic speed. The mass which we are observing is related to rest mass $m_0$ and the energy corresponds to peculiar velocities. However, in the models to calculate, for instance, critical density, we should take into account the increase of energy and in that way increase of total density of the universe due to the radial motion of the galaxies by the relativistic factor $\sqrt{1/(1-v^{2} /c^{2})}$. To account for the 96\% of the missing density, galaxies should have speed equal to 99.2\% of the speed of light, which is in agreement with our model. This is also in agreement with the theoretical predictions for the speed of thin shell expansion given in \cite{Czachor}.

One of the global methods for determining the mass density of the universe is by measuring the rate of cosmic expansion in the distant past. It is accepted that observations of type Ia supernovae at redshift $z$ $<$ 1 provide startling and puzzling evidence that the expansion of the universe at the present time appears to be accelerating, behavior attributed to ``dark energy'' with negative pressure \cite{6,7}.

This conclusion is based on the measurements of supernovae distances without using Hubble's law and their redshifts, the rate of the cosmic expansion in the distant past. According to the data, galaxies at large distances are receding less rapidly than Hubble's law would predict. A purely kinematic interpretation of the SN Ia sample provides evidence at the $>$ 99\% confidence level for a transition from deceleration to acceleration or similarly strong evidence for a cosmic jerk \cite{8}.

Our model provides an alternative explanation. It predicts with statistical significance that there is no acceleration. It clearly demonstrates that the deviation from Hubble's law is misinterpreted, and that it is not related to the accelerating expansion of the universe. The deviation is related to kinematics, to our position in the universe, and our point of sight.

To demonstrate this, let us stay consistent with our model and assume that galaxies are moving with a radial speed close to $c$, as explained in figure \ref{fig2}a. Let us again assume that our position in the universe is at point A. To measure the speed of another galaxy located on the line of sight from point A to point B, we will divide that arc into $n$ segments. Let us assume that one observer is located at each of the $n$ points on the segment from A to B. Each of these observers measures relative speed, $dv$, to its next neighbors (all will measure the same $dv$, since the segments have the same length). The observer at the point A will obtain the speed of the $m$-th observer by adding $m$ times differences $dv$ measured by each observer. Since the total sum will increase in value, the observer will need to apply the relativistic equation for the velocity addition. It is easy to show that a simple recurrence relation for the speed at segment $m$ has the form:
\begin{equation} \label{GrindEQ__12}
v_i= (v_{i-1}+dv)/(1+v_{i-1}dv/c^2),
\end{equation}
with $m$ iterations.

\begin{figure}
		\includegraphics[width=8cm]{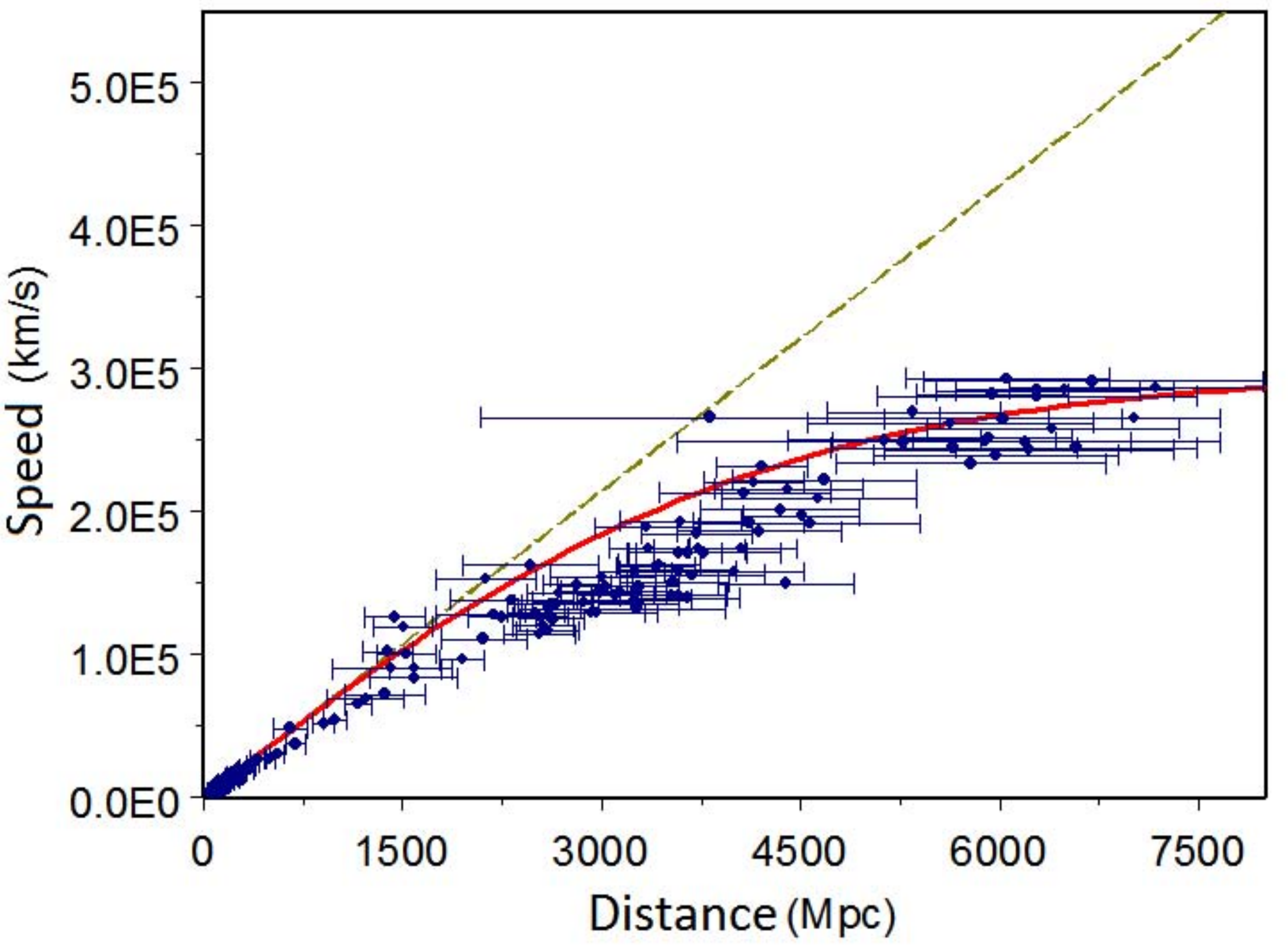}
		\caption{\label{fig3}
Hubble's velocity is represented by dashed line. The speed, as seen by an observer on the surface of the sphere, as a function of the distance on the sphere, is represented by the red curve. The experimental data from \cite{8} are plotted as points with assigned errors.
}
\end{figure}

The result of applying this equation is the curve plotted in Fig. \ref{fig3} which represents the speed (as seen by observer at point A) as a function of the distance (the length of the arc from the point A to the point of the measurement). Plotted on the same graph are Hubble's law $v(Hubble)=mdv$ and the experimental data from \cite{8}, where speed is obtained through redshift $z$ as $v=cz$, and luminosity distance $dL$ (in units of megaparsecs) is obtained from extinction-corrected distance moduli,
$\mu = 5 log dL + 25 $. Fig. \ref{fig3} clearly demonstrates that there is no acceleration of the universe, and that the effect is introduced by relativity. This result also underlines our earlier statement that there is no reason to introduce  dark energy, because the observations of the  acceleration of the universe can be explained by our model and kinematics.


Our model can be tested. Assuming that matter is homogenously distributed in the universe, a simple experiment which will count the number of galaxies as function of redshift could provide a test for space curvature. If space is in form of a shell, the number of galaxies as function of altitude on the sphere, or function of redshift, should first increase and then decrease. This test is more complex than it appears, since it should take into account the expansion of the space with time and the detection limits of current instrumentation, but it is feasible at the present time. The Hubble Ultra Deep Field (HUDF) optical and near-infrared survey performed in 2004 covered only a tiny patch of the sky, just 3.5 arc minutes across, but due to the high sensitivity and long exposure time extends thousands of megaparsecs away. HUDF shows a uniform distribution of matter by distance. This is consistent with the model, since integration by longitude could result in different number of galaxies for different redshifts, but a survey that will confirm this needs to be performed. However, it is important to note that a hollow shell model completely reproduces the distribution of the entire observed radio sources count for the flux density $S$ from $S\approx 10$ $\mu$Jy to $S\approx 10$ Jy \cite{Condon}.

\section{Conclusion}
 The gravitational field of the visible universe is so strong that it creates a photon sphere of radius $R_{ps} = 14.3$ Gpc. The model for the size of the visible universe must be corrected to account for the fact that measured horizon distance of 14.0 $\pm$ 0.2 Gpc is an arc of the circle rather than a straight line, because the entire visible universe is inside the photon sphere. The visible universe is a sphere (shell) with radius 4.46 $\pm$ 0.06 Gpc and event horizon is the maximal length of the arc on the surface (shell) of that sphere, which has the size of 14.0 $\pm$ 0.2 Gpc. Consistent with this definition is a model of the universe that assumes radial expansion of space with a radial speed $v_r$ close to $c$ and confinement of the galaxies and motion of light to the surface (shell) of the sphere $R_0=v_rt_0$. Such model gives an explanation for the Hubble law and predicts a value for the Hubble's constant of $H_0$ = 67.26 $\pm$ 0.90 km/s/Mpc and values for the cosmic expansion rate $H(z)$ that are in agreement with observations. It explains that the observed CMB originates from a single point on the opposite side of the sphere and that for that reason the measured CMB must be exactly the same for all directions of measurements, if corrected for fluctuations caused by large scale structures. It explains uniformity of the CMB without the inflation theory and may predict low variance temperature interference pattern. The model gives correct values for the particle horizon $\pi ct_0$ and the velocity of the particle horizon $2c$.
 The model predicts, by gravitation, correct values for the CMB redshift, $z=1166$, and time dilatation of Type Ia supernovae.
 It accounts for the missing density without introducing dark mass and dark energy, by applying relativistic energy correction. The model explains that type Ia supernovae redshifts are not related to the accelerated expansion of the universe, but are rather caused by our position in the universe and kinematics. In addition it explains that the reason for the established discrepancy between non-covariant version of the holographic principle and the calculated dimensionless entropy, $S/k$, for the visible universe that exceeds the entropy of a black hole is due to misinterpretation of the size of the visible universe. The model is in agreement with the distribution of radio sources in space, type Ia data, and with HUDF optical and near-infrared survey performed in 2004.

\begin{acknowledgments}
I would like to thank S. Matinyan and I. Filikhin for useful discussions. This work is supported by NSF award HRD-0833184 and NASA grant NNX09AV07A.
\end{acknowledgments}

\end{document}